\documentclass[twocolumn,showpacs,preprintnumbers,amsmath,amssymb]{revtex4}

\usepackage{graphicx}
\usepackage{dcolumn}
\usepackage{bm}
\begin{document}

\title{Phase diagram of optimal paths}


\author{Alex Hansen}
\email[Alex.Hansen@phys.ntnu.no]{}
\affiliation{Department of Physics, Norwegian University of Science and 
Technology, N-7491 Trondheim, Norway}


\author{J\'anos Kert\'esz}
\email[Kertesz@phy.bme.hu]{}
\affiliation{Institute of Physics, Budapest University of Technology,
Budafoki \'ut 8, H-1111 Budapest, Hungary} 


\date{\today}

\begin{abstract}
We show that choosing appropriate distributions of the randomness, the
search for optimal paths links diverse problems of disordered media
like directed percolation, invasion percolation, directed and
non-directed spanning polymers. We also introduce a simple and
efficient algorithm, which solves the $d$-dimensional model
numerically in ${\cal O}(N^{1+d_f/d})$ steps where $d_f$ is the
fractal dimension of the path. Using extensive simulations in two
dimensions we identify the phase boundaries of the directed polymer
universality class. A new strong-disorder phase occurs where the
optimum paths are self-affine with parameter-dependent scaling
exponents. Furthermore, the phase diagram contains directed and
non-directed percolation as well as the directed random walk models at
specific points and lines.

\end{abstract}


\maketitle

\begin{figure}
\includegraphics[width=7cm]{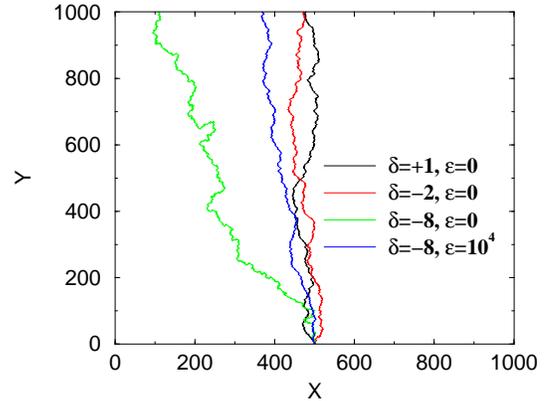}
\caption{Optimal paths for different values of the
disorder parameters $\delta$ and $\epsilon$, defined in Eq.\ (3).}
\label{fig1}
\end{figure}

The search for optimal paths in a random environment is one of the
basic problems of statistical physics \cite{Halpin}. The task is to
find the path between two points in a random energy landscape such
that the total energy along that path is minimal. After a mapping to a
graph where the edges bear random weights we arrive at a widely
studied problem of discrete mathematics (see, e.g.,
\cite{math}). From the point of view of physics, the interest comes from
intimate relations to various fields including the geometry of domain
walls in disordered magnets \cite{Huse,Kardar,Cieplak}, vortices in
superconductors \cite{Nelson} or rupture lines \cite{fracture}. For
$d$-dimensional {\it directed\/} lines without overhangs, often referred
to as ``directed polymers in random media'' (DPRM), there is an
exact mapping to stochastic surface growth \cite{KPZ}, enabling
further connections to fractal surfaces \cite{VF}, to driven particle
systems \cite{Derrida} and to the stochastic Burgers equation
\cite{Sinai}. The main questions for all these applications are:
How can the geometry of the optimal paths be described and how do the
energy fluctuations scale with their length?  Some realizations of optimal
paths for different disorders are shown in Fig.\ \ref{fig1}.

The situation is quite clear for the DPRMs: The line is a self-affine
fractal resulting in scaling endpoint-fluctuations. Imagine that the
polymer is held fixed at one end and we ask for conformations of
length $L$ over different disorder realizations. Let $(\Delta x)^2$ be
the average squared endpoint-fluctuations of the polymer, then
$\Delta x \propto L^\zeta$ where $\zeta $ is the roughness
exponent. Similarly, for the average squared energy fluctuations
$(\Delta E)^2$ we have $\Delta E
\propto L^\omega$. In $d=2$, the exponents are known exactly
\cite{Kardar}: $\zeta_{DPRM} = 2/3 $ and $\omega=1/3$. In higher dimensions
numerically calculated values are available \cite{forest}.

Recently, the question of the conformation of {\it spanning
non-directed\/} polymers has attracted considerable interest. Indeed,
the original optimal path problem does not exclude any geometry a
priori, in principle overhangs should be allowed. In fact, it can be
easily shown that overhangs have to occur on a sufficiently long line.
However, recent numerical work on such non-directed polymers in a
random medium (NDPRMs) have given evidence that, for bounded unimodal
distributions of disorder, these overhangs are irrelevant
\cite{stella} and the NDPRMs are in the universality class of DPRMs
\cite{havlin, rieger}. The latter conclusion was also drawn from a
real space renormalization group study \cite{stella}.

Most of the studies of optimum paths have been confined to Gaussian
distribution of the randomness.  The effect of changing the
distribution has been studied in some recent papers.  In spite of earlier
reports \cite{newman}, universality of the DPRMs was found for wide
family of bounded distributions \cite{perlsman}. It was shown \cite
{tang} that, when the distribution has a power law decaying part for
negative energies characterized by an exponent $\mu$, the effect
described by Zhang \cite{zhang} for surface growth can be observed for
DPRMs too: The scaling exponents of the paths depend continuously on
$\mu$ in an intervall of $\mu$.  For a bimodal distribution, where the
energy can be either $1$ or $0$ (with probability $p$ and $1-p$,
respectively) there are crossover phenomena near to the percolation
threshold $p_c$ \cite {stella, bimodal}.  Long range correlations in the
randomness may lead to nonuniversal behavior \cite {corr}.

The standard models of disorder, directed percolation (DP) or ordinary
percolation \cite {stauffer} can also be formulated in terms of global
optimization problems \cite {percopt}, as long as the spanning paths
are concerned. We briefly summarize these cases \cite{geometry}.  The
optimal path $\cal P$ in the polymer problem is given by
$$
\min_{\cal P} \sum_{i \in {\cal P}} E_i\;,
\eqno(1)$$
while in the percolation problems we have for the spanning paths $\cal P$
$$
\min_{\cal P} \left(\max_{i \in {\cal P}}\right) E_i\;,
\eqno(2)
$$
where for the directed versions the condition of directedness to (1)
and (2) have to be superimposed. The only difference between these
equations is that in (2) we have the maximum instead of the sum in
(1). When looking for the bottlenecks along each portion of the path
we obtain exactly an optimal percolation path. However, this path does
not scale like the shortest (or chemical) distance on a critical
percolation cluster, but rather as one on an invasion percolation (IP)
tree \cite {Cieplak, havlin_s}, or trapping invasion percolation
\cite{havlin_s}. These invasion percolation problems have much in
common with ordinary percolation, e.g., the critical point and the
fractal dimension of the critical cluster agree.

\begin{figure}
\includegraphics[width=6cm]{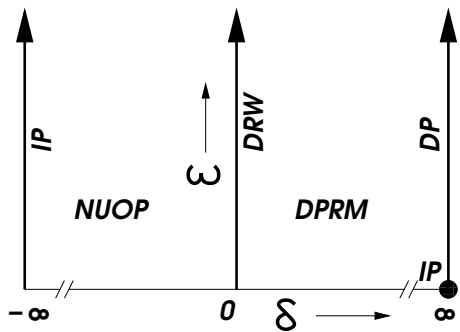}
\caption{Phase diagram in the disorder parameter $\delta$ and
the directedness parameter $\varepsilon$, defined in Eq.\ (3).
DPRM: Directed Polymer in a Random Medium, IP: Invasion Percolation,
DP: Directed Percolation, NUOP: Non-Universal Optimal Paths.
\label{fig2}}
\end{figure}

The geometry of the optimal percolation paths is also known. For DP we
have again a self-affine line with $\zeta_{DP}= 0.633$ in two
dimensions \cite {guttmann}, calculated from the ratio of the
perpendicular and the parallel correlation length exponents. For IP we
have a fractal line with a dimension $d_f > 1$. In two dimensions its
value is $d_f = 1.22$ \cite{Cieplak, havlin_s}. Naturally, the
roughness exponent for a fractal line is $\zeta _{IP}=1$.

It is tempting to try to formulate an unified picture of these
diverse but obviously related phenomena and in this Letter we address
this point. Furthermore we would like to explore the possibility of
new universality classes and at the same time to check earlier results
by improving the numerics.

Let us look for a parametrized distribution of randomness such that both
directedness and strong disorder naturally occur as limits. In order
to achieve this, weights are assigned to each bond as
$$ 
E = r^\delta + \varepsilon\;, 
\eqno(3)
$$ 
where $r \in (0,1)$ is taken
from an uniform random distribution; $\varepsilon \ge 0$ and $-\infty
\le \delta \le \infty$ are the directedness and the disorder
parameters, respectively. This choice corresponds to a PDF in the
reduced variable $\tilde E = E - \varepsilon$:
$$ 
p(\tilde E) = \frac{{\tilde E}^{1/\delta - 1}}{|\delta|}\;.
\eqno(4)
$$
Here $\tilde E \in (0,1]$ for $\delta > 0$ while $ \tilde E \ge 1$ 
for $\delta < 0.$ The optimal path is searched for in a random
environment characterized by the bond weights in (3), resulting in complex
phase diagram, Fig.\ \ref{fig2}.

Already a simple analysis of (3) reveals interesting connections
to different models of statistical physics. Let us take first
$\delta = 0$. Along this line a penalty is to be paid for any
overhang. However, all directed paths have the same weights. This is
exactly the statistics of directed random walks (DRW), leading to self
affine lines with roughness exponent $\zeta_{DRW} = 1/2$ \cite
{bikas}.  

The case of NDPRM with a bounded distribution can be identified with
$\varepsilon = 0$, $0 < \delta \le \infty$. The optimal paths are in the
universality class of DPRM \cite {havlin, rieger}. Obviously, this is
even more so when $\varepsilon $, the parameter suppressing overhangs, is
switched in.  Interestingly, no crossover phenomena at $\delta \to 0$
are to be expected since however small $\delta $ is the global
optimization over the random weights matters, since the $\epsilon $
weights represent only a constant background.

The limit $\delta \to +\infty$ the distribution is more and more
shifted towards a delta function at 0. Nevertheless, this limit corresponds 
to {\it strong
disorder} as it can be shown with reference to the order statistics
\cite{d81}: The cumulative distribution is $P(E)=\int_0^E p(E')dE'
=E^{1/\delta}$ (we
first assume $\varepsilon=0$). We generate an ensemble of groups of
$E$-values, $N$ in each group.  We order the numbers in each group in
an ascending order so that $E_{(1)}\le E_{(2)}\le \cdots \le E_{(N)}$.
The average value of the $n$th element in this ordering is $\langle
E_{(n)}\rangle =[n/(N+1)]^\delta$.  Hence, the ratio between
consequtive elements in this ordering is given by $\langle
E_{(n+1)}\rangle/ \langle E_{(n)}\rangle =(1+1/n)^\delta$ which
diverges for any $n$ as $\delta\to +\infty$.  Consequently, any sum of
elements drawn from this distribution will be dominated by the largest
element, showning that we are in the infinite disorder limit, and the
optimum paths are percolation paths \cite {percopt}. For $\varepsilon
= 0$ it is the invasion percolation shortest path \cite
{havlin_s}. However small value of $\varepsilon $ is taken it will
immediately suppress the overhangs transforming the optimal path to a
critical {\it directed\/} percolation path.

\begin{figure}
\includegraphics[width=6cm]{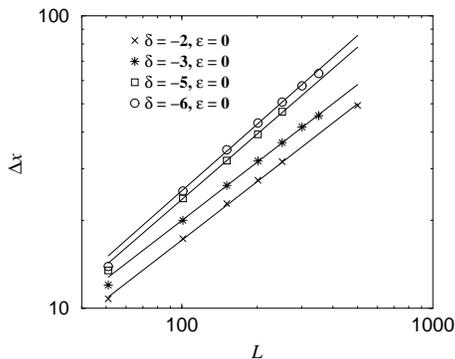}
\caption{$\Delta x$ as a function of system size $L$.  The slopes of
the $(\delta=-2,\varepsilon=0)$ and $(-3,0)$ lines
are 0.67, while the slopes of the $(\delta=-5,\epsilon=0)$ and $(-6,0)$ lines are 0.75.
\label{fig3}}
\end{figure}

In the regime $\delta < 0$ the distribution has a power law tail
with the consequence that only the $k < 1/|\delta|$-th moments remain
finite. If this property is inherited to the optimal paths, this would
mean that there should be nonuniversal behavior as a function of
$\delta$ since the universality classes are characterized not only by the
exponents but also by the universal scaling functions. Moreover, the
large number of huge obstacles generated by the tail of the distribution
may lead to an enhancement of overhangs resulting in new geomtries.
We apply numerical tools in order to see clearly in this
point.

Before that, let us summarize the behavior at $\delta \to -\infty$.
In this limit, the distribution becomes not normalizable indicating the
dominance of large values of $E$, i.e., this corresponds again to the infinite
disorder limit, where the arguments of order statistics apply
straitforwardly and we arrive at (2) which is the percolation case. It
is a question of interest whether the paramter $\varepsilon$ is
relevant for this limit and we will come later back to this point.

Let us turn to our two-dimensional numerical simulations. The obvious
difficulty in the determination of optimal paths is in the global
nature of the problem. Given an optimal path of span $L$, the solution
for size $L+1$ can be entirely different. However, there are
algorithms which calculate the optimal path on graph with random
positiv weights on the edges in polynomial time \cite {math}.  We have
developed an efficient algorithm which is adequate to the geometry we
considered.

We now describe this algorithm in some detail.  It is closely related
to the algorithm described by Hansen and Hinrichsen in \cite {percopt}
for determining the non-directed percolation threshold. We describe
the algorithm for the two-dimensional case, i.e., for a $45^\circ$
tilted square lattice. The sample has a cylindrical geometry (periodic
boundary conditions in the horizontal direction). Each bond $i$ is
assigned an energy $E_i$.  We associate a variable $V_{\alpha}$ with
each node $\alpha$.  Initially, these variables are all set to zero
with the exception of the nodes forming the horizontal edges of the
lattice.  These nodes are assigned a value which is larger than any
energy $E_i$ appearing in the system.  Furthermore, the node forming
the anchor-point of the polymer, $\alpha=0$ is assigned a negative
value $V_0=v_0=-|v_0|$.  This node sits on the lower edge of the
lattice.  We then iteratevely update the {\it internal\/} vertices
(i.e.\ those not being on the horizontal edges of the lattice)
according to the scheme
\begin{eqnarray*}
\label{update}
V_{\alpha} \to \min(V_{\alpha(1)}+E_{\alpha(1)},
V_{\alpha(2)}+E_{\alpha(2)},\nonumber\\
V_{\alpha(3)}+E_{\alpha(3)},
V_{\alpha(4)}+E_{\alpha(4)})\;,
\end{eqnarray*}
where $\alpha(1)$ to $\alpha(4)$ are the four nodes that neighbor node
$\alpha$ and the four bonds joining them.  The updating proceeds until
the node variables no longer change.  The value of each node variable
is then equal to the energy of the optimal path between that
particular node and the anchor node plus the value $v_0$. The number
of iterations goes as $L \times $ the length of the shortest path,
i.e., as $L^{1+d_f}$, where $d_f$ is the fractal dimension of that path.

\begin{figure}
\includegraphics[width=6cm]{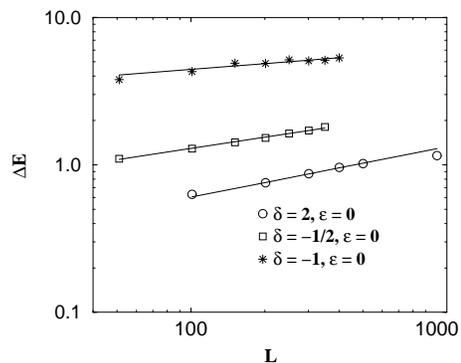}
\caption{$\Delta E$ as a function of system size $L$.  
The slopes are for $(\delta=2$,$\varepsilon=0$)
0.33, for $(-1/2,0)$, 0.25 and for $(-1,0)$ 0.13.
\label{fig4}}
\end{figure}

In Fig.\ 3 we show end point fluctuations $\Delta x$ as a function of
system size $L$ for a number of different disorders.  For $\delta \ge -2$
and $\varepsilon=0$,
we find $\Delta x \sim L^{0.67}$, indicating self affinity with the same
roughness exponent as in the DPRM problem, where $\zeta_{DPRM}=2/3$.  We
also find the same behavior with $\delta=-8$ and $\varepsilon=10^4$.  However,
for disorders with $\delta < -5$ or less and $\varepsilon=0$, we find
$\Delta x \sim L^{0.75}$, which indicates self affine behavior, but with a
new roughness exponent which is significantly larger than the DPRM one.
The self affinity of the optimal paths is further supported by monitoring
the average length of the paths, $l$ as a function of system size $L$.  For
all finite disorders investigated, we find $l\sim L^{1.00}$.  Hence, the curve is
{\it not\/} a self-similar fractal, and the overhangs are irrelevant.  Fig.\ 1 shows
different optimal paths with different disorders.  Even in the most extreme
cases ($\delta=-8$ and $\varepsilon=0$), there are few overhangs.

The fluctuations in energy, $\Delta E$, scales as $\Delta E \sim L^{1/3}$
in the DPRM case.  For $\delta > -1$ we find that $\Delta E$
is self averaging.  In Fig.\ 4, we show $\Delta E$ as a function of $L$
for $\delta=2$, $\delta=-1/2$, and $\delta=-1$, all with $\varepsilon=0$.  
For $\delta=2$ case, we find an exponent consistent with the DPRM value, 1/3.
However, for the $\delta=-1/2$, we find a value which is much smaller, 0.25,
while for $\delta=1$, we find 0.13. This may be a crossover effect.

For values of $\delta <  -1$, $\Delta E$ is not self averaging, indicating
no universal scaling functions exist.  However, the scaling of $\Delta x$ with
two different values of scaling exponent suggests that there are still
partial universality in the system for small values of $\delta$ and $\varepsilon=0$,
with a roughness exponent of 0.75.

Universality classes are characterized by scaling functions, not only
by exponents \cite{HHmoments}. Therefore we extended our simulations
to higher moments of the endpoint fluctuations (wherever it was
possible), because the amplitude ratios are known with high precision
\cite{HHmoments} and also to the study of the moments of the energy
fluctuations. From our studies we conclude: i) For $0 < \delta <
\infty$ the amplitude ratios are within the numerical accuracy the
same as for DPRM, confirming universality. ii) For $0 > \delta >
-\infty$ the $k$-th moments of the energy fluctuations become
divergent for $k>1/|\delta|$ (i.e., even the second moment is
nonexistent for $\delta < 2$) indicating {\it nonuniversal\/} behavior
(see. Fig.\ \ref{fig2}: Non-universal optimal paths, NUOP). The
measurements were carried out for $\varepsilon = 0$. In the case of
$\varepsilon > 0$ the situation is even worse: While $\varepsilon$
does not contribute to the energy fluctuations, its effect is enhanced
directedness, i.e., the path is forced through higher $E$ valued
regions. However, it seems that unversality in a restricted sense is
present: The roughness exponent seems to agree with $\zeta _{DPRM}$
for $\delta \ge -4$ and $\varepsilon =0$. Also for large values of
$\varepsilon$ the $\zeta $ becomes close to $2/3$ even for large
negative values of $\delta$. Furthernmore, the roughness exponent for
$\delta < -4$ is $\approx 0.75$. Further investigations are needed to
explore fully this restricted universality.

We thank K.\ Sneppen and S.\ Havlin for useful comments.  Support was
provided by the NFR through grant SUP/10225700. JK is member of the
Center for Applied Mathematics and Computational Physics at the BUTE.

\end{document}